\documentclass{aa}

\usepackage{natbib}
\bibpunct{(}{)}{;}{a}{}{,} % to follow the A&A style

\usepackage{graphicx}
\usepackage[small]{caption}

\begin{document}

\title{Evidence for TeV gamma ray emission from Cassiopeia A}

\titlerunning{Evidence for TeV gamma ray emission from Cassiopeia A}

%\thesaurus{09.19.2; 09.09.1 Cas\,A; 13.07.2; 09.03.2}

\author{
F.~Aharonian\inst{1},
A.~Akhperjanian\inst{7},
J.~Barrio\inst{3},
K.~Bernl\"ohr\inst{1},
H.~B\"orst\inst{5},
H.~Bojahr\inst{6},
O.~Bolz\inst{1},
J.~Contreras\inst{3},
J.~Cortina\inst{2},
S.~Denninghoff\inst{2},
V.~Fonseca\inst{3},
J.~Gonzalez\inst{3},
N.~G\"otting\inst{4},
G.~Heinzelmann\inst{4},
G.~Hermann\inst{1},
A.~Heusler\inst{1},
W.~Hofmann\inst{1},
D.~Horns\inst{4},
A.~Ibarra\inst{3},
C.~Iserlohe\inst{6},
I.~Jung\inst{1},
R.~Kankanyan\inst{1},
M.~Kestel\inst{2},
J.~Kettler\inst{1},
A.~Kohnle\inst{1},
A.~Konopelko\inst{1},
H.~Kornmeyer\inst{2},
D.~Kranich\inst{2},
H.~Krawczynski\inst{1,9},
H.~Lampeitl\inst{1},
M.~Lopez\inst{3},
E.~Lorenz\inst{2},
F.~Lucarelli\inst{3},
N.~Magnussen\inst{6,10},
O.~Mang\inst{5},
H.~Meyer\inst{6},
R.~Mirzoyan\inst{2},
A.~Moralejo\inst{3},
E.~Ona\inst{3},
L.~Padilla\inst{3},
M.~Panter\inst{1},
R.~Plaga\inst{2},
A.~Plyasheshnikov\inst{1,8},
J.~Prahl\inst{4},
G.~P\"uhlhofer\inst{1},
G.~Rauterberg\inst{5},
A.~R\"ohring\inst{4},
W.~Rhode\inst{6},
G.~P.~Rowell\inst{1},
V.~Sahakian\inst{7},
M.~Samorski\inst{5},
M.~Schilling\inst{5},
F.~Schr\"oder\inst{6},
M.~Siems\inst{5},
W.~Stamm\inst{5},
M.~Tluczykont\inst{4},
H.J.~V\"olk\inst{1},
C.~A.~Wiedner\inst{1},
W.~Wittek\inst{2}}

\institute{Max-Planck-Institut f\"ur Kernphysik, Postfach 103980, D-69029 Heidelberg, Germany
\and Max-Planck-Institut f\"ur Physik, F\"ohringer Ring 6, D-80805 M\"unchen, Germany
\and Universidad Complutense, Facultad de Ciencias F\'{\i}sicas, Ciudad Universitaria, E-28040 Madrid, Spain
\and Universit\"at Hamburg, II. Institut f\"ur Experimentalphysik, Luruper Chaussee 149, D-22761 Hamburg, Germany
\and Universit\"at Kiel, Institut f\"ur Experimentelle und Angewandte Physik, Leibnizstra{\ss}e 15-19, D-24118 Kiel, Germany
\and Universit\"at Wuppertal, Fachbereich Physik, Gau{\ss}str.20, D-42097 Wuppertal, Germany
\and Yerevan Physics Institute, Alikhanian Br. 2, 375036 Yerevan, Armenia
\and On leave from Altai State University, Dimitrov Street 66, 656099 Barnaul, Russia
\and Now at Yale University, P.O. Box 208101, New Haven, CT 06520-8101, USA
\and Now at IFAE, Unversitat Aut\`onoma de Barcelona, Spain
}

\authorrunning{Aharonian et al.}

\date{Received 23 November 2000 / Accepted 23 January 2001}

\offprints{\\G.~P\"uhlhofer,
\email{Gerd.Puehlhofer@mpi-hd.mpg.de}}

\abstract{232 hours of data were accumulated from 1997 to 1999, using the 
HEGRA Stereoscopic Cherenkov Telescope System to observe
the supernova remnant Cassiopeia A. TeV $\gamma$-ray
emission was detected at the $5\,\sigma$ level, and a flux of 
$(5.8 \pm 1.2_{\mathrm{stat}} \pm 1.2_{\mathrm{syst}}) \times 10^{-9}
\mbox{ph}\,\mbox{m}^{-2} \mbox{s}^{-1}$ above 1\,TeV was derived.
The spectral distribution is consistent with a power law with a differential spectral 
index of $-2.5 \pm 0.4_{\mathrm{stat}} \pm 0.1_{\mathrm{syst}}$ between 1 and 10\,TeV.
As this is the first report of the detection of a TeV $\gamma$-ray source on the ``centi-Crab'' scale,
we present the analysis in some detail. Implications for the acceleration of cosmic rays
depend on the details of the source modeling. 
We discuss some important aspects in this paper.
\keywords{ISM: supernova remnants -- ISM: individual objects: Cassiopeia A -- cosmic rays -- gamma rays:
observations}
}

\maketitle

%
%________________________________________________________________

\section{Introduction}
Supernova remnants (SNRs) are widely believed to be the acceleration sites for cosmic rays (CR)
 -- ions as well as electrons -- 
up to particle energies of at least $10^{15}\,\mbox{eV}$ \citep[see][ for recent reviews]
{VoelkKruger,BaringSnowbird}.
Astronomical evidence for this theory,
i.e. via identification of sources, is based on the detection of accompanying $\gamma$-rays which are
produced in or nearby the source.

Hadronic CR produce $\gamma$-rays in collisions with gas, mainly via $\pi^{0}$-decay. 
At GeV energies, SNR source candidates compete with
the diffuse Galactic $\gamma$-ray background \citep{EgretSnrs}. 
Optimum sensitivity is currently expected at TeV energies \citep{DAV};
nevertheless, a clean detection is lacking. 

Hard synchrotron X-ray spectra of several SNRs, 
including \object{SN\,1006} \citep{AscaSN1006},
\object{SNR\,RX\,J1713.7-3946} \citep{AscaSN1713},
and \object{Cassiopeia A} (Cas\,A) \citep{AllenCasA,LetterSAXCasA},
have been interpreted 
as evidence for non-thermally accelerated electrons up to $\approx 100\,\mbox{TeV}$. 
Also, TeV $\gamma$-rays are generated from electron populations by bremsstrahlung 
and by inverse Compton (IC) upscattering of ambient soft photons, e.g. from the microwave background.
Observations of TeV photon emission from SN\,1006 and SNR\,RX\,J1713.7-3946,
reported by the CANGAROO experiment \citep{CangarooSN1006,CangarooSN1713},
have been interpreted in this framework, although a hadronic origin should not be excluded
\citep{AharonianSN1006,BerezhkoICRC}.

Cas\,A is the brightest shell-type SNR that is accessible for the
HEGRA experiment. ``Brightest'' applies to the radio band 
\cite[][ and references therein]{Atoyan:CasA1}
as well as to non-thermal X-rays \citep{AllenPCA}.
The remnant results from the
youngest known Galactic supernova which dates from around 1680.
Its distance is estimated at 3.4\,kpc \citep{ReedCasA}.
The images clearly reveal the shell-type nature of the remnant. 
Only recent high resolution X-ray
maps from the Chandra satellite \citep{ChandraCasA} indicate a central object.
The progenitor of Cas\,A was probably a Wolf-Rayet star, as discussed in 
\citet{FesenCasAProgenitor} and \citet{ComptelCasATi44}.
Its initial mass is estimated to be $\approx 30\,\mbox{M}_{\sun}$. The supernova blast wave is
expanding into a wind bubble and shell system from the previous wind phases of the progenitor star,
which plays an important role in the modeling of the shock acceleration
of CR in such remnants \citep{BerezhkoBubbles}. 
At TeV energies, upper limits have been 
given by Whipple \citep{WhippleSNRs} and CAT
\citep{CatCasA}.
Nevertheless, detectable flux levels can be expected both
from electron emission and the hadronic channel, as we will discuss in 
\S\,\ref{implications}.

In this paper, we report the results obtained with the HEGRA Stereoscopic Cherenkov Telescope System.
A preliminary analysis, based on parts of the data, was presented in 
\citet{ICRCHEGRACasA} and already 
revealed evidence for TeV $\gamma$-ray emission. 
A summary of the observations and analysis results is given in the next section.
\S\,\ref{anadetails} deals in more detail with the analysis procedure.
Finally, in \S\,\ref{implications} we interpret our results
in the context of model predictions for the TeV $\gamma$-emission
from Cas\,A.

%
%________________________________________________________________

\section{Observations and summary of the analysis results}
\label{observations}

The HEGRA Stereoscopic Cherenkov Telescope System \citep{HEGRAPerformance99} is located on the
Roque de los Muchachos on the Canary Island La Palma, at 2200\,m above sea level. It
consists of 5 identical telescopes (CT\,2 - CT\,6), which operate in coincidence for
stereoscopic detection of air showers induced by primary $\gamma$-rays in the atmosphere.

Observations were performed in the summer months of 1997, 98 and 99. Cas\,A can be observed 
from the HEGRA site at zenith angles of 29\degr~or larger.
The average zenith angle was 32\degr; observations typically did not go beyond 40\degr.
Hence, the data set provides a $\gamma$-shower peak detection energy of $1\,\mbox{TeV}$
for \object{Crab}-like spectra \citep{HEGRAPerformance99}.
Data were taken with 3, 4, and 5 active telescopes. Data cleaning
consists of bad weather rejection and exclusion of telescopes with technical problems.
Data from the telescope CT\,2, which was included last into the system
and was still being tuned, were excluded from this analysis.
The cleaned data set contains 232 hours of observations.

Shower images from a telescope are accepted in the analysis if they contain at least
40 photoelectrons.
The particle direction is determined using a stereoscopic reconstruction algorithm described in
\citet{HofmannStereoTechniques}.
The data set contains $2.1\cdot 10^{6}, 2.5\cdot 10^{6}, \mbox{and}\,2.9\cdot 10^{6}$ events,
for 2, 3, and 4 views available in an event, respectively.
Events are accepted within a circle of 1\degr\,radius from the center
of the field of viev (FOV). 
Without filtering, virtually all events 
-- coming from the source direction as well as from the surrounding sky region --
are induced by hadronic CR's.
Candidates for $\gamma$-rays are selected against this background
using a cut on the image shape parameter {\it mean scaled width} ($msw$) \citep{Mrk501PaperI}
of $0.5<msw<1.1$.

\begin{figure}
  \centering
  \includegraphics[width=7.5cm]{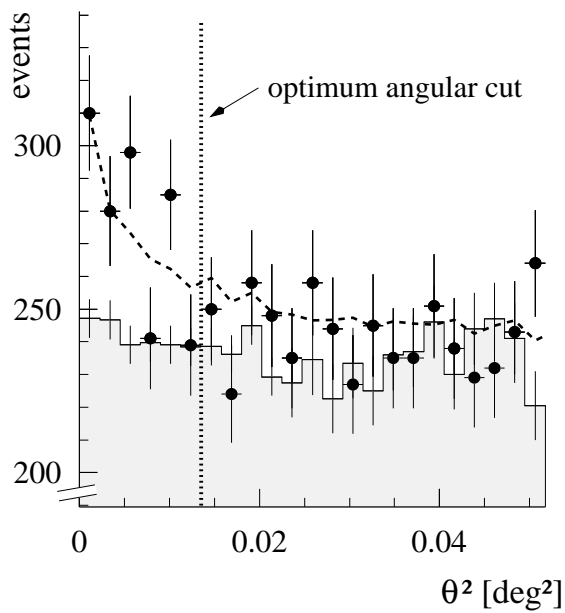}
  \caption[]{Dots: Number of events vs. the squared angular distance
  to the position of Cas\,A.
  Shaded histogram: Background estimate; 
  up to $\theta^{2}=0.0225\,\degr^{2}$, data from 7 control regions is used.
  Therefore, the statistical error of the background estimate is much smaller than the
  error of the source distribution.
  The dashed line shows Crab excess events, measured at similar zenith angles, 
  scaled down to 3.3\%,
  and superimposed on a flat background.
  The vertical dotted line indicates the position of the optimum angular cut.}
  \label{casaexcess}
\end{figure}

In Fig.~\ref{casaexcess}, the number of events is plotted vs. the angular distance
to the position of Cas\,A. 
The background which remains after the shape cut is estimated using seven control regions
in the FOV.
The optimum angular cut is derived from Crab and \object{Mrk\,501} data.
The excess significance, calculated after \citet{LiMa}, is $4.9\,\sigma$ for this straightforward
evaluation. 

The photon flux and energy spectrum of Cas\,A is derived by comparison with a large Crab data sample
\citep{KonopelkoCrab}, taken between 1997 and 2000.
For the spectral analysis, we use the energy reconstruction method described in 
\citet{Mrk501PaperI}.
The spectral distribution is comparable with the distribution measured for the
Crab nebula. 
Under the assumption of a power law spectrum 
$\mbox{d}F_{\mathrm{\gamma}} / \mbox{d}E \propto E^{\alpha}$ from 1 to 10\,TeV,
we derive a differential spectral index of 
$\alpha = -2.5 \pm 0.4_{\mathrm{stat}} \pm 0.1_{\mathrm{syst}}$.
The flux is 3.3\% of the Crab flux, this corresponds to 
$F(E>1\,\mbox{TeV})= (5.8 \pm 1.2_{\mathrm{stat}} \pm 1.2_{\mathrm{syst}})
\times 10^{-9}\,\mbox{ph}\,\mbox{m}^{-2}\,\mbox{s}^{-1}$.

%
%________________________________________________________________

\section{Analysis details}
\label{anadetails}

In this paper, we present for the first time evidence for a TeV signal from a source 
on the ``centi-Crab'' flux scale, which was obtained after a very deep observation over three years.
Therefore, some analysis details will be explained here.
%Since we present evidence for the first time in this paper for a TeV signal from a source
%on the ``centi-Crab'' flux scale, which was obtained after
%very deep observation over three years, some analysis details will be explained here.

The data sample is background-dominated, even after strong background suppression cuts.
Control of systematic effects in background
determination is therefore far more important than in the case of strong TeV $\gamma$-ray sources,
like the Crab nebula \citep{KonopelkoCrab} or Mrk\,501 \citep{Mrk501PaperI};
\S\,\ref{background} deals with the relevant issues.
Strong background suppression cuts are needed to obtain a significant signal from Cas\,A;
\S\,\ref{shapecuts} and 
\S\,\ref{angres} describe the corresponding procedures.
The performance of the HEGRA system changed over the observation time, and in
\S\,\ref{performance} the necessary calibration is discussed.
For the determination of the flux and the
spectrum of Cas\,A, we apply the same calibration and cut procedure
to a Crab reference data sample, and compare both $\gamma$-ray samples 
(\S\,\ref{fluxandspectrum}).
In this analysis, Cas\,A is treated as a TeV $\gamma$-ray point source;
\S\,\ref{sourcepos} deals with the topic of source localization and extension.

\subsection{Background estimation}
\label{background}

\begin{figure}
  \resizebox{\hsize}{!}{\includegraphics{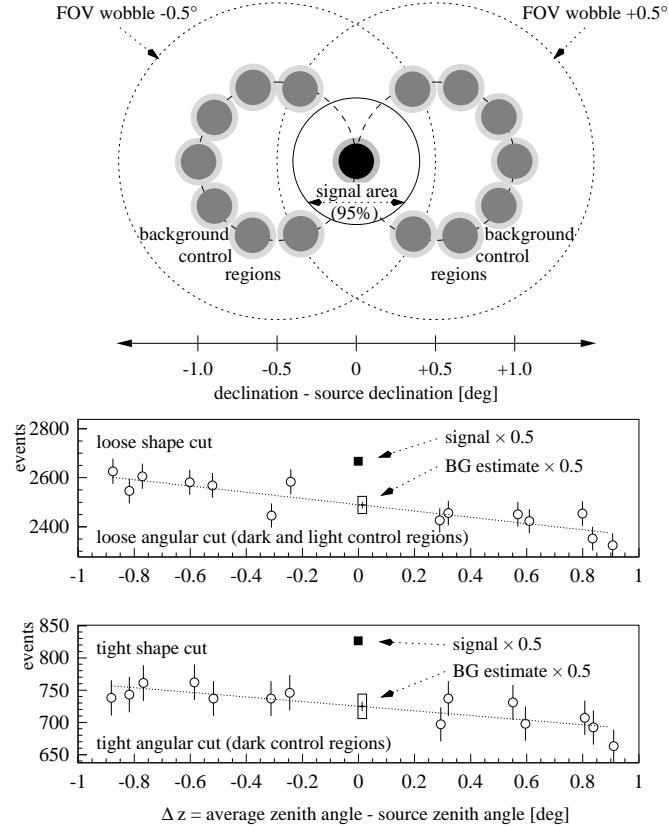}}
  \caption[]{Upper panel: Background control region setup.
             The dotted circles indicate the FOV, which is alternatingly shifted by plus or minus $0\fdg5$ in declination
	     relative to the source position. The solid circle contains 95\% of the signal in case of a point source.
             Middle and lower panels: Events per control region, as a function of the average zenith angle
	     in the control region with respect to the source zenith angle. 
             The shape cuts are defined in \S\,\ref{shapecuts}. 
	     In the middle panel,
     	     a loose shape cut and a loose angular cut of $\theta < 0\fdg15$ are used. 
	     In the lower panel, a tight shape cut and a tight angular cut of $\theta < 0\fdg12$ are applied.
             Each open circle corresponds to one background control region, 
	     either in wobble $-0\fdg5$ ($\Delta z < 0\degr$)
	     or $+0\fdg5$ mode ($\Delta z > 0\degr$).
             As the background estimate for the source region, the average over all control regions is used;
	     the value is shown as the small centered cross.
             The open box indicates the expected $1\,\sigma$ background fluctuation range. 
             The filled square shows the number of events in the source region,
             normalized by a factor 0.5, since both the wobble $+0\fdg5$ and $-0\fdg5$ observing
	     modes contribute to the signal.
	     The statistical significance for the excess in the lower panel is $4.9\,\sigma$.}
\label{controlreg}
\end{figure}

The HEGRA system reveals a nearly flat acceptance for shower directions
over the FOV of $2\degr$ diameter,
while 95\% of the signal of a point source is contained in 
a region with a diameter of $0\fdg 8$ 
around the source\footnote{This is essentially true 
also in the case of slightly extended sources such as possibly Cas\,A,
see also \S\,\ref{sourcepos}}.
Therefore we observe
point sources in the so-called wobble mode, where the source is displaced $\pm0\fdg 5$ in
declination from the center of the FOV, the sign being reversed every 20 or 30 minutes.
The background can hence be estimated from simultaneously gathered data, and systematic effects 
in the background estimation due
to weather or performance changes essentially cancel out.

Variations of background rates across the FOV are caused by slight acceptance changes,
see e.g. \citet{ICRCGalplane}, and possibly by differences in
night sky light noise.
For the analysis here, we use seven non-overlapping control regions 
centered on a circle of radius $0\fdg 5$ around the center of the FOV.
The setup is sketched in Fig.~\ref{controlreg}, upper panel.
This method has the following characteristics:\\
(a) It provides better statistics than in the standard case using only one opposite region;
    the background control area is seven times larger than the signal area.\\
(b) The regions view different sky areas, closer and further away from the
    source, and can be cross-checked for sky noise influence
    (see Fig.~\ref{controlreg}, middle and lower panels).\\
(c) The change in radial acceptance
    is fully taken into account.\\
(d) Any constant gradient in the background distribution is compensated for;
    the most significant contribution is due to a zenith angle dependence of the event rate, as
    can be seen in Fig.~\ref{controlreg}, middle and lower panels.\\

As the background estimate, the average over all control regions is used.
A systematic error in the background estimation
can possibly be caused by higher-order FOV inhomogeneities.
The background levels of the control regions which are closest
to the source position deviate by less than 2\% from that estimated
for the source location;
we consider therefore the systematic error of the background estimation to be below 2\%.

In the $\theta^{2}$-distribution (Fig.~\ref{casaexcess}), the background estimation is produced
by a properly normalized overlay of all control regions. Note however that the setup described
above only works at $0 \le \theta^{2} \le 0.0225\,\degr^{2}$.
Above $0.0225\,\degr^{2}$, the background is estimated from adjacent control regions
(not shown in Fig.~\ref{controlreg}),
which (I) have only twice the area than the signal region,
and (II) cannot fully take the radial acceptance change into account. However,
this exclusively affects the right part of Fig.~\ref{casaexcess}, but no other
aspect of this analysis.

\subsection{Shape cuts}
\label{shapecuts}

\begin{figure}
  \resizebox{\hsize}{!}{\includegraphics{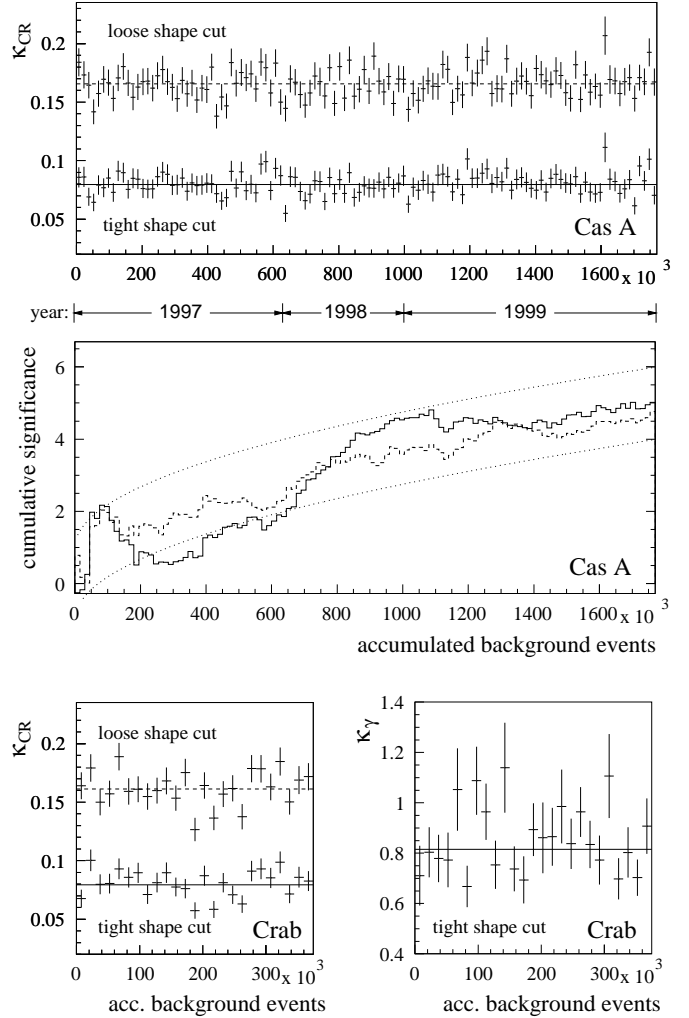}}
  \caption[]{
             Upper panel: 
             For the Cas\,A data set, the survival probability of background events
             is plotted, broken down into observation time slices; for the ``time'' axis, 
             the number of accumulated background events in an enlarged FOV of $\diameter=2.8\degr$ is used.
             The values are derived from 
             the off source control regions, applying tight and loose shape cuts.
             Middle panel:
             The evolution of the Cas\,A signal with the total number of background
             events is shown. 
             For a constant signal, the significance should develop as
             [background events]$^{-1/2}$ with a variation of $1\,\sigma$; the dotted
             lines indicate this expected range, assuming a $5\,\sigma$ signal at the
             end of the observation.
             Lower left panel:
             For the Crab data set, the same values as in the upper panel are shown;
             this demonstrates the good agreement of the Cas\,A and the
	     Crab reference data.
             Lower right panel:
             In addition, the $\gamma$ efficiency of the tight shape cut is shown for
             the Crab data set. The values are derived by comparing the background
             subtracted sample after the tight shape cut with the background
             subtracted sample after a cut of $0.5<msw<1.4$,
             which rejects virtually no $\gamma$-rays.
             For all figures, an angular cut of $0\fdg12$ was applied.
            }
  \label{scaledwidth}
\end{figure}

A cut on the image parameter {\it mean scaled width} ($msw$) is applied to
reduce the background induced by charged CRs \citep{Mrk501PaperI}.
The {\it scaled width} is the measured width of an event seen by one telescope, relative to the
expected value for a $\gamma$-shower with the given measured distance and image amplitude.
By averaging over all telescopes included in an event, one obtains the {\it mean scaled width}.
The $\gamma$-ray distribution peaks at 1; its RMS width is approx. 0.1.
Charged CR, on the other hand, produce more diffuse images,
hence the background distribution peaks (with a much broader
distribution) at 1.4--1.55; the exact value changes with zenith angle.  
The $\gamma$-ray expectation values are usually determined by shower and detector simulations
\citep{HEGRAPerformance99}.

Analyses which focus on the reduction of systematic influences
of the background suppression on the $\gamma$-ray efficiency $\kappa_{\mathrm{\gamma}}$, use a loose cut of
$0.5<msw<1.2$ ($\kappa_{\mathrm{\gamma}} = 95\%$), e.g. \citet{Mrk501PaperI}.
In the search for faint sources, optimum sensitivity
is obtained by a tight cut of $0.5<msw<1.1$,
with an expected $\gamma$-ray efficiency of $\kappa_{\mathrm{\gamma}}=80\%$. 
Since this cut is more dependant on the exact location of the $\gamma$-ray peak
in the {\it mean scaled width} distribution, the expectation values were
refined to match the quasi background-free $\gamma$-ray distribution of the Mrk\,501 data taken in
1997 \citep{Mrk501PaperI}. The $\gamma$-ray {\it mean scaled width} distribution is thereby
shifted by 3.5\%, to be now exactly centered at 1. This was verified with our
calibration source, the Crab nebula.

In addition, since Crab is not permanently observable, we use the survival
probability for background events (typically $\kappa_{\mathrm{BG}}=8\%$
for the tight, and $16\%$ for the loose shape cut,
at the given zenith angle range) as an indicator for the stability of $\kappa_{\mathrm{\gamma}}$.
From background control regions in the FOV,
$\kappa_{\mathrm{BG}}$ can be continuously derived (see Fig.~\ref{scaledwidth} upper panel).

In the context of this analysis, the loose shape cut is only used as an additional
consistency check. We note that even if $\kappa_{\mathrm{\gamma}}$ changes, this 
would only induce an error at the flux determination;
excess significances are not affected due to the
simultaneous measurement of background events.

\subsection{Angular resolution}
\label{angres}

\begin{table}[t]
  \begin{center}
%  \begin{tabular}{l|l|l|l|l|l} \hline \hline
  \begin{tabular}{llllll} \hline \hline
%      \multicolumn{6}{|l|}{Li \& Ma, equation 17}              \\ \hline
      \multicolumn{6}{l}{Li \& Ma, equation 17}              \\ \hline
      & \# off            & $\alpha$       & \# on           & \# $\gamma$ rays          & significance \\ \hline
    A & 10152             & 1/7            & 1653            & 203                       & 4.9          \\ \hline
    B & 3165              & 1/7            & 579             & 127                       & 5.3          \\ \hline \hline
%      \multicolumn{6}{|l|}{full likelihood analysis}                                                    \\ \hline
      \multicolumn{6}{l}{full likelihood analysis}                                                    \\ \hline
%      & \multicolumn{3}{|l|}{\# complete FOV}                & \# $\gamma$ rays          & significance \\ \hline
      & \multicolumn{3}{l}{\# complete FOV}                & \# $\gamma$ rays          & significance \\ \hline
%    C & \multicolumn{3}{|l|}{103270}                         & 190                       & 5.6$^{\dag}$ \\ \hline \hline
    C & \multicolumn{3}{l}{103270}                         & 190                       & 5.6$^{\dag}$ \\ \hline \hline
  \end{tabular}
  \end{center}
  \caption[]{ Excess significances for different cuts and signal evaluations.
              All values are derived after tight shape cuts.
              The number of events is referred to by the symbol \#.
              Lines A and B are evaluated with different angular cuts (see text for details);
              $\alpha$ is the signal to background area normalisation factor
              ($\alpha=A_{\mathrm{on}}/A_{\mathrm{off}}$).
              Line C corresponds to the full likelihood analysis introduced in the text.
              \dag: preliminary value.}
  \label{eventnumbers}
\end{table}

The angular resolution changes with zenith angle and shape cut event selection.
The applied stereo reconstruction algorithm 
(described by \citet{HofmannStereoTechniques}, algorithm 2)
allows the prediction of the resolution on an event basis
from event parameters with an accuracy of 15\%.

The standard analysis does not make use of this information; one
simply accepts events up to a maximum distance from the source.
The given data sample yields a median $\gamma$-ray resolution of $0\fdg 09$;
the optimum cut here is $0\fdg12$, which was derived from the 
point sources Crab and Mrk\,501.
This cut rejects 40\% of the $\gamma$-rays.
For Cas\,A, the excess significance, following \citet{LiMa}, amounts to $4.9\,\sigma$
(see Table \ref{eventnumbers} line A).

In a simple approach to make use of the known angular resolution on an event basis,
events were divided according to the number of telescopes participating in an
event (4, 3, 2 telescopes).
The three classes differ significantly both in the angular resolution and the background rejection.
Different angular cuts were derived for all classes simultaneously.
In order to optimize the sensitivity for the given data sample, 
the angular cuts were derived such that a data sample which contains 5\% of the
Crab $\gamma$-ray flux yields the optimum significance; this training was
performed on real data from
Crab and Mrk\,501\footnote{
After tight cuts, signal and background levels start to be at the same order of magnitude 
(see e.g. Fig.~\ref{casaexcess}). Hence further optimization
of the so-called Q-factor $=\kappa_{\mathrm{\gamma}}/\sqrt{\kappa_{\mathrm{BG}}}$ does not yield 
the optimum sensitivity
($\kappa_{\mathrm{\gamma}}$ and $\kappa_{\mathrm{BG}}$ are the cut survival probabilities for $\gamma$'s
and background events, respectively).}.
The resulting cuts are $0\fdg11$ for 4-telescope events, $0\fdg08$ for 3-telescope events,
and the complete rejection of events with only 2 participating telescopes.
The excess significance is again derived by a simple counting
of signal and background events \citep{LiMa},
and amounts to $5.3\,\sigma$ (Table \ref{eventnumbers} line B), in agreement with
the expected sensitivity improvement derived from Crab data.

In order to fully exploit the stereo information, a maximum likelihood analysis was performed,
see e.g. \citet{AlexandreasPointSourceSearch} and references therein.
All events after tight shape cuts in the FOV of $\diameter=2\degr$ are used.
The probability density function (PDF) for signal events $P_{\mathrm{s}}$ uses
the event-wise predicted 2-dimensional error matrix 
$\tens{C}$ for the angular resolution, 
determined according to the prescription in \citet{HofmannStereoTechniques}:
\begin{displaymath}
  P_{\mathrm{s}}(\tens{C};\vec{x}) = \frac{1}{2 \pi \sqrt{\left| \tens{C} \right|}} \: \exp\left(-\frac{1}{2}\:
                                     \vec{x}^{T} \tens{C}^{-1} \vec{x}\right),
\end{displaymath}
where $\vec{x}$ denotes the FOV vector of the reconstructed event direction.
The PDF for background events $P_{\mathrm{b}}$ is calculated with an acceptance function
which parametrizes the acceptance of the complete FOV as a function of radius $r$
and zenith angle component $z$ of $\vec{x}$:
\begin{displaymath}
  P_{\mathrm{b}}(\vec{x}) = P_{\mathrm{b}}(r,z) ,\hspace{1ex} r = \left| \vec{x} \right| ,\hspace{1ex} z = \vec{x} \cdot
                            \vec{e}_{\mathrm{zenith}}.
\end{displaymath}
The likelihood function $\cal L$ is given by
\begin{displaymath}
  {\cal L}(N_{\mathrm{s}}) = {\textstyle \prod} \left[ r_{i} N_{\mathrm{s}} P_{\mathrm{s}} + \left( N - r_{i}
                             N_{\mathrm{s}} \right) P_{\mathrm{b}} \right].
\end{displaymath}
Here, $N$ is the total number of events, and $N_{\mathrm{s}}$ is the number of $\gamma$ candidates coming from the source.
We also use an a-priori expected ratio for 4-, 3-, and 2-telescope signal events $r_{i}$ 
($r_{i=4,3,2} \approx 0.40:0.35:0.25$, changing with zenith angle); the values were derived from
Crab and Mrk\,501 $\gamma$-data.
For Cas\,A, the likelihood ratio test yields a significance of $5.6\,\sigma$ (Table \ref{eventnumbers} line C).
Our likelihood analysis is still under investigation.
The influence of different background parametrizations, obtained from different data sets, was
found to be small but not negligible.

\subsection{Performance changes}
\label{performance}

The HEGRA system has shown performance changes over the past years which are attributed
mainly to the following three effects:\\
(1) Mirror misalignment, which temporarily lead to changes in the shape cut efficiencies. 
In this analysis, only Crab data were affected, and data of these periods were rejected.
The stability of $\kappa_{\mathrm{\gamma}}$ is monitored as discussed in the previous paragraph
(see Fig.~\ref{scaledwidth}, upper and lower panels).\\
(2) Aging of mirrors. \\
(3) Aging of photomultipliers (PMs);
the PM gain can be determined independently using laser calibration runs as described in
\citet{MarkusThesis}. In spring of 1999, the high voltage of all PMs was increased to 
compensate for the previous PM gain loss of 15\%.

The effects (2) and (3) lead to a change of the light sensitivity of the telescope system.
Due to the resulting change in energy threshold, the global sensitivity change can be monitored
by the CR trigger rate. The system nominally runs at a trigger rate of 15\,Hz 
for a 4-telescope configuration, but the rate temporarily dropped to 70\%.
Both factors are entered into the calibration procedure to provide an adjusted energy scale.

For the spectral evaluation of the signal, an extended software threshold is applied,
which yields a constant energy acceptance (and also event rate) over time.
The extended software threshold emulates the hardware trigger threshold of the telescopes,
but on the recalibrated energy scale. It was increased until the resulting CR event rate was 
constant over time. Thus, data of Cas\,A and Crab, which were taken at different observation periods,
can be directly compared. Uncertainties in the determination of the energy acceptance function
(so called collection area, e.g. \citet{HEGRAPerformance99}), which may arise due to the
tight angular and shape cuts, and due to the changing energy threshold, are avoided.

Of course, this method only works at the expense of number of events.
However, the Crab $\gamma$-ray and the background rates show a constant ratio, even without this 
extended software threshold. For the flux evaluation, it can therefore be omitted in case of
Crab-like spectra, in order to have the full statistics available. For compensation of the
different detection rates, Cas\,A and Crab reference data were normalized to each other
using total CR event rates.

In Fig.~\ref{scaledwidth} middle panel we show the development of the Cas\,A signal
as a function of the effective exposure time, as determined by the background event rate.
Within statistical errors, the excess significance does not deviate from the behaviour
which is expected for a steady signal.

\subsection{Flux and energy spectrum}
\label{fluxandspectrum} 

\begin{figure}
  \resizebox{\hsize}{!}{\includegraphics{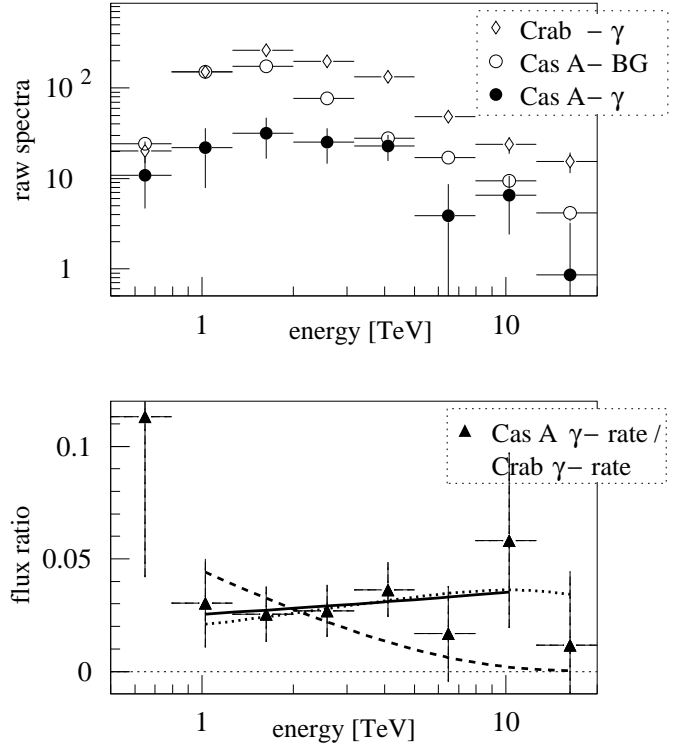}}
  \caption[]{
  Differential spectra of the Cas\,A and Crab data sets.
  In the upper panel, raw spectra are plotted: 
  ($\diamond$) Crab on source, background subtracted;
  ($\bullet$) Cas\,A on source, background subtracted;
  ($\circ$) expected background of the Cas\,A data sample, which is derived from the background
  control regions and scaled to the source area. 
  The background level of the Crab data set is approx. 5 times lower, due to the
  shorter exposure time of this reference set. The lower panel shows the
  ratio of the Cas\,A and Crab $\gamma$-ray spectra, both normalized by the CR rate
  (i.e. essentially by the observation time). The solid line shows the fit of a power law
  $\mbox{dF}_{\gamma} / \mbox{dE} \propto \mbox{E}^{\Delta\alpha}$ with 
  $\Delta\alpha = \alpha_{\mbox{\scriptsize Cas\,A}} - \alpha_{\mbox{\scriptsize Crab}}$.
  The dotted and dashed curves are fits to the spectral shapes of the 
  hadronic and leptonic emission models which are discussed in \S\,\ref{implications}.} 
  \label{casaspectrum}
\end{figure}

For flux and spectral evaluation, we compare the data from Cas\,A with the ``standard candle''
Crab nebula. The HEGRA Crab data have been evaluated in great detail to derive the Crab
flux and spectrum, using data from the full zenith angle range up to $60\degr$
\citep{KonopelkoCrab,KonopelkoLargeZenithAngle}.
As reference, we use the Crab flux of  
$(\mbox{d}F_{\mathrm{\gamma}} / \mbox{d}E)_{\mathrm{Crab}}
 = (2.79 \pm 0.02 \pm 0.5) \cdot 10^{-7} 
 (\frac{E}{1\,\mathrm{TeV}})^{-2.59 \pm 0.03 \pm 0.05}\mbox{ph}~\mbox{m}^{-2}
\mbox{s}^{-1}\mbox{TeV}^{-1}$ between 1 and 20\,TeV.

For the determination of the spectral distribution, we apply the energy reconstruction method
described in \citet{Mrk501PaperI}, which provides an energy resolution of 20\%.
The extended software threshold as introduced in \S\,\ref{performance}
is used, in order to obtain adjusted Cas\,A and Crab reference data samples.
The peak detection energy of $\gamma$-rays is derived from Crab data
and amounts to 1.4\,TeV for this analysis.

Figure~\ref{casaspectrum} upper panel shows the $\gamma$-ray spectra of Cas\,A and Crab,
as well as the background distribution of the Cas\,A data sample. In Fig.~\ref{casaspectrum},
lower panel, the $\gamma$-ray spectrum of Cas\,A is divided by the Crab
$\gamma$-ray spectrum. In the spectral analysis, data below 1\,TeV are rejected, since 
acceptance errors due to the slight difference in the zenith angle
distribution of the Cas\,A and the Crab reference data samples become important.

The fit to a power law spectrum was a priori restricted to the energy range from 1 to 10\,TeV:
since the predicted spectra for different model assumptions (see \S\,\ref{implications})
show cutoffs at higher energies,
only a local power law is a reasonable assumption in the comparison between data and predictions;
on the other hand, the statistics do not allow
much further restriction of the energy range.
Assuming for Cas\,A a power law spectrum $\mbox{d}F_{\mathrm{\gamma}} / \mbox{d}E \propto E^{\alpha}$,
we derive a differential spectral index of 
$\alpha_{\mathrm{Cas\,A}} = -2.5 \pm 0.4_{\mathrm{stat}} \pm 0.1_{\mathrm{syst}}$
between 1 and 10\,TeV (see solid line in Fig.~\ref{casaspectrum}, lower panel).

We also compare the measured spectrum of Cas\,A to the predicted spectral shapes
for hadronic or leptonic emission, as described in \S\,\ref{implications}.
In this test, all data above 1\,TeV are included. The fits are shown as the dotted line
for the hadronic spectrum, and as the dashed line for the leptonic spectrum. 
In order to test the compatibility of the data with either model,
two statistical tests were applied, a simple $\chi^{2}$-test and a Kolmogorov test \citep{Eadie}.
The latter is more powerful, since it is also sensitive to a sequence of consecutive deviations of the same sign.
While the hadronic spectrum has a slightly better chance probability 
(93\% in the $\chi^{2}$-test, 98\% in the Kolmogorov test),
the leptonic spectrum has a chance probability of 31\% in the $\chi^{2}$-test and 15\%
in the Kolmogorov test and is therefore not ruled out.

We note that a pure background sample would lead to a spectrum which is steeper 
by $\Delta \alpha \approx -1$. This can be taken as additional evidence that
the observed signal from the direction of Cas\,A is not due to underestimation of the expected background\footnote{
Note that the background is a strongly $\gamma$-ray selected sample, and evaluated with a $\gamma$-ray
energy reconstruction algorithm. Therefore the background energy spectrum does not represent the charged CR
spectrum, but rather the spectrum of $\gamma$-ray-like events (fluctuations of
charged CR's, and possibly diffuse $\gamma$-rays and electrons).}.

For Cas\,A's TeV $\gamma$-ray flux, one obtains 3.3\% of the Crab flux; results with and without
the extended software threshold are in good agreement.
We derive a flux of 
$F_{\mathrm{Cas\,A}}(E>1\,\mbox{TeV})= (5.8 \pm 1.2_{\mathrm{stat}} \pm 1.2_{\mathrm{syst}})
\cdot 10^{-9}\,\mbox{ph}\,\mbox{m}^{-2}\,\mbox{s}^{-1}$.

\subsection{Source position}
\label{sourcepos} 

\begin{figure}
  \resizebox{\hsize}{!}{\includegraphics{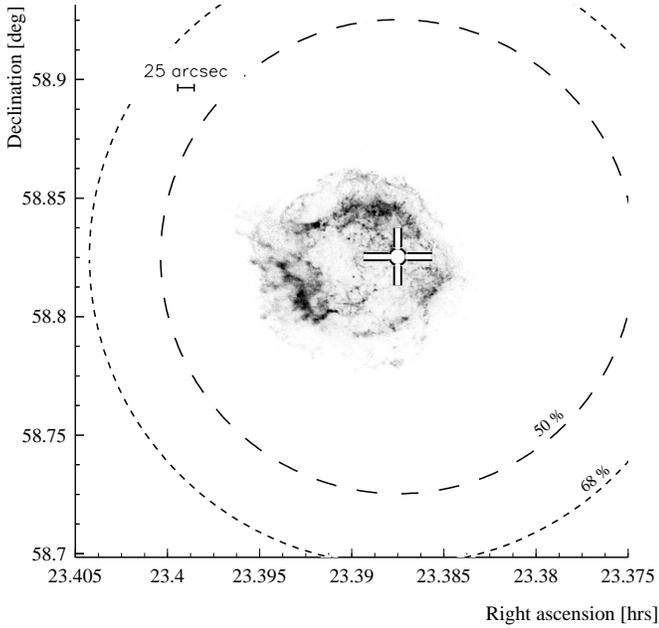}}
  \caption[]{Cas\,A 
             (Ra 23\,h 23\,min 24\,s, Dec $58\degr 48\farcm 9$)
             in year 2000 celestial coordinates. The cross shows the
             reconstructed position of Cas\,A from the TeV $\gamma$-ray excess,
	     the error bars give the $1\,\sigma$ statistical error of the reconstruction.
             The dashed and dotted circles indicate the 50\% and 68\% single event error, respectively.
             The systematic pointing uncertainty of the system is 25 arcsec.
             To indicate the scale, the plot is superimposed on the recent Chandra X-ray image 
	     (courtesy of NASA/CXC/SAO).}
  \label{casaposition}
\end{figure}

Cas\,A's outer shock has a diameter of 5\arcmin, which is slightly below the RMS angular resolution
of the HEGRA telescope system. For the Crab nebula, which has similar extensions, recent studies have
shown that resolving this scale is marginally out of reach for the system,
even with high statistics and a low zenith angle $\gamma$-ray sample \citep{HofmannCrabSize}.

For Cas\,A, the strongest spatial deviation from a point source at the center of the remnant
could be caused by a hot spot at the shell, as may be indicated by spectrally and spatially resolved
X-ray images \citep{BeppoSAXCasA,ASCACasA,ChandraCasA}.
%PSPC Rosat image \citep{Aschenbach}.
Even in this case, the optimum angular cut for a point source would reject
less than 5\% more $\gamma$'s than expected.
Therefore, deviating from the preliminary analysis presented in 
\citet{ICRCHEGRACasA}, we treat Cas\,A as a point source.  

Figure~\ref{casaposition} shows the position of Cas\,A, reconstructed from TeV $\gamma$-rays,
and superimposed on the high resolution Chandra X-ray image \citep{ChandraCasA}.
Pointing calibration and source reconstruction uses
the methods described by \citet{HEGRAPointing}.
The source position is determined by a fit of a 2-dimensional Gaussian to
the measured events in celestial coordinates.
In order to increase the signal-to-noise ratio, 
only 3- and 4-telescope events of the full data sample were used (see
\S\,\ref{angres}).
We conclude that within statistical errors,
the TeV emission is centered on the source, and extended
emission can neither be proven nor be excluded.

%
%________________________________________________________________

\section{Implications for cosmic ray acceleration}
\label{implications}

The detection of TeV $\gamma$-rays proves that Cas\,A is a site of CR acceleration for 
particles -- either nucleons or electrons -- with multi-TeV energies.
Since it is a shell-type remnant, this detection adds further support 
to the theory of SNRs being responsible for CR acceleration
via the shock acceleration mechanism. However, further conclusions rely on the
identification of the hadronic and/or leptonic nature of the high energy primary particles.

Considerable effort has gone into the understanding of the primary electron population, which can be
traced by its synchrotron radiation at various wavebands, from radio to X-rays. Especially the hard
X-ray spectrum has been interpreted as stand-alone proof for 40 TeV electrons \citep{AllenCasA}.
A source model has been developed to predict the TeV emission in the framework of a multi-wavelength
study \citep{Atoyan:CasA1}.
The basic challenge is the electron transport in the highly non-homogenous source.
The variations in the prediction for TeV emission via the inverse Compton (IC) mechanism,
which will dominate the emission above 1\,TeV, are mainly due to different deduced electron spectra;
the target photon density -- the synchrotron photons, the thermal dust emission in the far infrared,
the optical/IR line photons, and the microwave background radiation -- is well-known.
X-ray emission, on the other hand, traces both the electrons and the magnetic field strength, which 
is typically derived to be of the order of milli-Gauss. The high energy X-ray tail 
and the TeV flux are spatially unresolved, and hence could be emitted in different regions
with different electron spectra.

The electron transport model which attempts to take the spatial structure of Cas\,A into
account predicts a range of possible fluxes \citep{Atoyan:CasA2}.
Interestingly, the shape of the spectrum at
TeV energies remains constant, and shows a steep cutoff with a
differential spectral index of below $-3$ above 1\,TeV.
If this holds true for all possible scenarios, it offers a chance to discriminate leptonically
induced $\gamma$-rays from the hadronic channel.

The acceleration of hadronic CR in the shock of a SNR is theoretically fairly well-understood for the
case of expansion into a homogenous interstellar medium 
\citep{BerezhkoUniform1,BerezhkoUniform2,RowellTycho}.
Strong modifications are expected if the shock expands into the wind structure,
e.g. of a Wolf-Rayet progenitor star \citep{BerezhkoBubbles}.
These calculations show that high $\pi^{0}$-decay $\gamma$-ray fluxes can be expected
even for young SNRs like Cas\,A. However, the influence of the magnetic field
configuration in progenitor winds may result in largely perpendicular shocks, with
strongly reduced injection efficiency,
which would lower the expected $\gamma$-ray fluxes considerably \citep{VoelkKruger}.

Given these arguments, the expected absolute flux level of $\pi^{0}$-decay $\gamma$-rays
is presently not well determined. The value assumed in \citet{Atoyan:CasA2}
and used later in the comparison
has been chosen such that the total energy of relativistic protons
is 20\% of the CR energy sum of about $10^{50}\,\mbox{ergs}$, which is ultimately released by an average Galactic SNR.
This reasonable choice does not violate the upper limits from the EGRET detector as well as from Whipple
\citep{WhippleSNRs} and CAT \citep{CatCasA} at TeV energies.
Since it cannot be increased substantially without doing so,
it can be considered as a model upper limit.
Again, the shape of the spectrum remains as a common feature, the $\pi^{0}$-decay spectrum
should extend up to 1 TeV with a hard $E^{-2.1..2.2}$ power-law.

\begin{figure}
  \resizebox{\hsize}{!}{\includegraphics{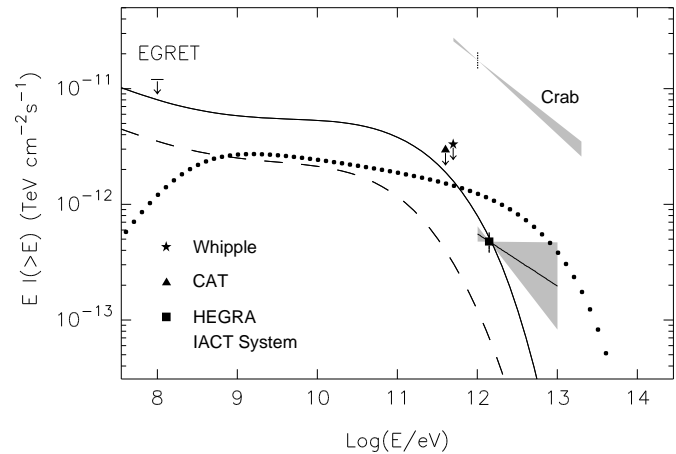}}
  \caption[]{The measured flux and spectral index of Cas\,A in the context of model predictions.
  The shaded area shows the $1\sigma$ error range for the measured spectral
  distribution under the assumption of a $\mbox{E}^{-\alpha}$ power law spectrum.
  The dotted curve 
  represents the
  model 
  spectrum for the $\pi^{0}$-decay flux as discussed in the text;
  the current model however allows a renormalisation of the spectrum.
  The solid and dashed lines show
  the predicted IC 
  plus bremsstrahlung flux 
  for different 
  model parameters; see \citet{Atoyan:CasA2} for details. 
  Also indicated are the upper limits measured by EGRET, Whipple and CAT.} 
  \label{casaflux}
\end{figure}

Figure~\ref{casaflux} shows the measured TeV flux of Cas\,A, together with the $1\,\sigma$ error
band of the spectral index under the assumption of a power-law spectrum between 1 and 10\,TeV.
The TeV upper limits are taken from \citet{WhippleSNRs} and \citet{CatCasA}.
Indicated are the expected spectra for IC plus bremsstrahlung emission, for different 
model parameters (solid and dashed lines) as described in \citet{Atoyan:CasA2}.
The dotted curve shows the predicted spectrum for the $\pi^{0}$-decay $\gamma$-ray flux.
\\
{\it Slope:}
          The slope of the spectrum marginally favours hadronic emission, but the data are
          also well compatible with the leptonic spectrum; the respective chance probabilities are
          98\% and 15\% (for details see \S\,\ref{fluxandspectrum}).
          Moreover, the discrepancies between data and leptonic models might also be attributable
          to the complexity of the source and the resulting possible model variations.
\\
{\it Integral flux:}
          Given the uncertainties in the acceleration efficiency and the resulting 
          uncertainty of the absolute $\pi^{0}$-decay flux as discussed above,
          the measured flux is consistent with hadronic emission;
	  it is also in good agreement with one leptonic prediction.

In summary, our results may at present be interpreted as 
leptonic or hadronic emission (or a mixture of both).
It remains to be seen whether the results of future theoretical work
using detailed acceleration models will allow definite
conclusions about the nature of the primary CR particles which lead to the TeV emission from Cas\,A.

\begin{acknowledgements}
The support of the German ministry for Research and
technology BMBF and of the Spanish Research Council CICYT is gratefully
acknowledged. G.~P.~R.\ acknowledges receipt of a Humboldt fellowship.
We thank the Instituto de Astrof\'{\i}sica de Canarias
for the use of the site and for supplying excellent working conditions at
La Palma. We gratefully acknowledge the technical support staff of the
Heidelberg, Kiel, Munich, and Yerevan Institutes.
\end{acknowledgements}

\bibliographystyle{apj}
\bibliography{10501}

\listofobjects

\end{document}